\documentclass[english]{revtex4-1}
\usepackage[T1]{fontenc}
\usepackage[latin9]{inputenc}
\usepackage{textcomp}
\usepackage{amsmath}
\usepackage{amssymb}
\usepackage{esint}

\makeatletter

\newcommand*\LyXZeroWidthSpace{\hspace{0pt}}

\usepackage[colorlinks = true,
            linkcolor = red,
            urlcolor  = blue,
            citecolor = blue,
            anchorcolor = blue]{hyperref}

\makeatother

\usepackage{babel}
\begin{document}

\title[NC Continuity Eq with Nl Potential]{{\normalsize{}ON THE FISK-TAIT EQUATION FOR SPIN-3/2 FERMIONS INTERACTING
WITH AN EXTERNAL MAGNETIC FIELD IN NONCOMMUTATIVE SPACE-TIME}}

\author{{\normalsize{}Ilyas Haouam}}

\email{ilyashaouam@live.fr}

\address{Laboratoire de Physique Mathématique et de Physique Subatomique (LPMPS),
Université Frères Mentouri, Constantine 25000, Algeria}
\begin{abstract}
{\normalsize{}In this paper, we investigated the Fisk\textendash Tait
equation in interaction with an external magnetic field in noncommutative
space-time. Knowing that the space-time noncommutativity is introduced
through the Moyal\textendash Weyl product known method. Consequently,
we studied the continuity equation in both commutative and noncommutative
space-time; there we examined the influence of the space-time noncommutativity
on the current density quadri-vector. Moreover, we find that the total
charge obtained from the probability density still indefinite even
when space does not commute. Furthermore, we found the spin current
density in the two different spin directions. We also investigated
the linking between the fermions and the bosons in the Fock space
using the Holstein\textendash Primakoff transformation.}{\normalsize \par}

$\phantom{}$

$\phantom{}$

\end{abstract}


\maketitle

\section{{\normalsize{}Introduction}}

Over the years, particle equations for an arbitrary spin was considered
the subject for careful investigations. That is why today we are interested
in the relativistic equations that describe the motion of spin-3/2
particles. Such as the relativistic Rarita\textendash Schwinger equation
(1940) \cite{key-1}, the Fisk\textendash Tait equation (1973) \cite{key-2},
Hurley's field equation \cite{key-3}, Bhabha\textendash Gupta equation
(Bhabha, Gupta 1952, 1954, 1974) \cite{key-4,key-5}, the approach
for arbitrary spin equation by V. Bargman, E.P. Wigner (1948) \cite{key-6}.
Also the Heisenberg equations of motion for the spin-3/2 field (1977)
\cite{key-7}, in which, it is shown for dynamical systems with constraints
depending upon external fields, the Lagrange and Heisenberg equations
of motion are the same for the quantized charged spin-3/2 field in
the presence of a minimal external electromagnetic interaction. The
Rarita\textendash Schwinger spin-3/2 equation in the weak-field limit
is obtained to satisfy the Heisenberg equations of motion. This is
similar to the case of spin-3/2 field minimally coupled with an external
electromagnetic field by Mainland and Sudarshan (1973) \cite{key-8}. 

As well recently, we have the link between the relativistic canonical
quantum mechanics of arbitrary spin and the covariant local field
theory by V.M. Simulik (2017) \cite{key-9} (where the found equations
are without redundant components). Where it has been confirmed that,
the synthesis of the relativistic canonical quantum mechanics of the
spins-3/2 particle and antiparticle doublet is completely similar
to the synthesis of the Dirac equation from the relativistic canonical
quantum mechanics of the spin-1/2 particle-antiparticle doublet. On
the basis of the investigation of solutions and transformation properties
with respect to the Poincare group the obtained new 8-component equation
is suggested to be well defined for the description of spin s=3/2
fermions. Despite the Rarita-Schwinger equation which has 16 components
and needs the additional condition.

The equation for particles of spin-3/2 originally was given by Fierz
and Pauli (I939) in spinor form \cite{key-10}. Knowing that the Klein-Gordon,
Dirac, and Proca equations provide a relativistic description of the
particles that have the lowest spins cases (S=0, 1/2, 1 respectively).

For instance, the Rarita\textendash Schwinger equation was formulated
for the first time by William Rarita and Julian Schwinger, it was
the most famous equation that describes the motion of the spin-3/2
fermions. However, in this research, we are interested in the Fisk\textendash Tait
equation. Because Rarita\textendash Schwinger equation and some of
the mentioned equations at the top of the introduction containing
many problems, such as the problem of causality, where the wavefront
propagation described by the corresponding equation is greater and
faster than that of the light \cite{key-11}. Also the problem of
the imaginary energy. Unlike the mentioned equations, with Fisk\textendash Tait's
equation all of these problems almost have been solved. To be clear
from the beginning all of the equations that describe the motion of
spin-3/2 fermions have the problem of indefiniteness of the total
charge \cite{key-12,key-13} including some exceptions such as in
the case of the Gupta equation \cite{key-14} the charge density is
positive-definite for some properties. As well as in the theory of
interacting spin-3/2 particle by T. Fukuyama and K. Yamamoto \cite{key-15},
where they suppress redundant particles by making their masses infinite.
But these choices make the total charge in the free theory still indefinite. 

Among the particles that have spin-3/2, we mention the gravitino,
the quasi-spin-3/2 particles from the pairs of Cooper, and the baryon.
This latter is a composite subatomic particle, which is a combination
of at least three quarks such as $u$ (up quark), $d$ (down quark)
and $s$ (strange quark) forming a baryon decuplet with spin-3/2,
as in the case of baryons which composed of one type of quark ($uuu$,
$ddd$, ...), these can exist in J = \LyXZeroWidthSpace 3\textfractionsolidus 2
configuration. But in J = \LyXZeroWidthSpace 1\textfractionsolidus 2
is forbidden by the Pauli exclusion principle. In the case of baryons
which composed of three types of quarks ($uds$, $udc$, ...), these
can exist in both J = \LyXZeroWidthSpace 1\textfractionsolidus 2 and
J = \LyXZeroWidthSpace 3\textfractionsolidus 2 configurations. Taking
into account that the first omega baryon discovered experimentally
was the $\varOmega^{-}$ hyperon, made of three strange quarks, in
1964 \cite{key-16}, then recently, in 2017 the LHCb collaboration
announced the observation of five new narrow $\varOmega_{c}^{0}$
states decaying to $\Xi_{c}^{+}K^{-}$ (an exceptionally large group
of baryons) \cite{key-17}.

The baryons (strongly interacting fermions) are acted on by the strong
nuclear force, are described by Fermi-Dirac statistics, which apply
to all particles obeying the Pauli-exclusion principle. Therefore,
the two most studied groups of baryons are $S=1/2$; $L=0$ and $S=3/2$;
$L=0$, which correspond to $J=1/2$ and $J=3/2$, respectively. Although
they are not the only ones, it is also possible to obtain $J=3/2$
particles from $S=1/2$ and $L=2$, as well as $S=3/2$ and $L=2$
\cite{key-18}. Theoretically, the baryons can have a higher spin
of 3/2.

The first known superconductor in which the quasi-spin-3/2 particles
form Cooper pairs was created by American and New Zealand physicists
\cite{key-19}. The un-conventional superconductor is an alloy of
Yttrium, Platinum, and Bismuth, which is normally a topological semi-metal,
the study was carried out by Johnpierre Paglione and his colleagues
at the University of Maryland, Ames Laboratory of the Iowa State,
Lawrence Berkeley national laboratory, and the universities of Otago
and Wisconsin. In the alloy studied by Paglione and his colleagues,
the charge is carried by quasi-particles of spin-3/2 particle type.
These quasi-particles result from interactions between the electron
spins and the positive charges of the atoms that compose the alloy,
this influence is called the spin-orbit coupling and it is especially
powerful in this material. Knowing that the spin-3 state, which combines
the moments of rotation and the orbital angular momentum, is the lowest
energy state.

The gravitino \cite{key-20,key-21} is the superpartner of the graviton,
predicted by the theories of the supergravity (this gauge fermion
mediating supergravity interactions, and its existence is only a hypothesis).
This spin-3/2 fermion obeys one of those mentioned equation at the
top of the introduction, and in super\textendash symmetry theory ($SUSY$)
it is assumed that each fundamental particle has a superpartner. For
a fermion (a particle with half-integer spin) such as the electron,
its superpartner would be a particle with an integer spin, it is the
selectron. For an integer spin particle such as the photon, its superpartner
would be a half-integer particle called photino. For the quantum of
gravitational force, in $SUSY$ when the particle carrier is the graviton
with spin-2, its superpartner is called the gravitino with spin-3/2,
such theory is called the Gauged Super Gravity theory.

For many years, there has been interesting in the study of spin-3/2
particles in as much detail as has been done for the spin-1/2 particles,
so that, our work considered as a contribution that may help in the
study of the spin-3/2 particles domain.

The goal of this work is to focus on the Fisk\textendash Tait equation
in a noncommutative space-time. Thereafter, extracting the continuity
equations in both commutative and noncommutative space-time through
the classical method, not using Lagrangian to extract the current
density. Then studying some additional contributions such as the current
density of the spin, also about how can link the fermions with bosons
in Fock space using the Holstein\textendash Primakoff representation.

\section{{\normalsize{}THE SPIN-3/2 EQUATION IN AN EXTERNAL MAGNETIC FIELD
IN NONCOMMUTATIVE SPACE-TIME}}

\subsection{{\normalsize{}THE FISK\textendash TAIT EQUATION }}

For the concept of the product, we shall always sum over repeated
indices. For example $A^{\mu}B_{\mu}=g^{\mu\nu}A_{\nu}B_{\mu}=A_{0}B_{0}-\overrightarrow{A}\overrightarrow{B}$,
with Greek letters $\mu$, $\nu$, $\lambda$ being $0,1,2,3$ and
$g^{\mu\nu}=diag\left(1,-1,-1,-1\right)$ is the metric tensor in
Minkowski space-time. We set $\hbar=c=1$(natural units).

Recently, Fisk and Tait proposed an equation for spin-3/2 particles
in 1973 \cite{key-22}, and they have been shown that their equation
remains causal even with minimal electromagnetic coupling, and made
it free of other types of difficulties by which the Rarita\textendash Schwinger
formalism is plagued. 

The wave function employed is a 24-component antisymmetric tensor-spinor
$\Psi_{\lambda}^{\mu\nu}=-\Psi_{\lambda}^{\nu\mu}$. It transforms
according to the following Lorentz group representation 
\begin{equation}
D(0,\frac{3}{2})\oplus D(\frac{3}{2},0)\oplus D(\frac{1}{2},1)\oplus D(1,\frac{1}{2})\oplus D(0,\frac{1}{2})\oplus D(\frac{1}{2},0),\label{eq:1}
\end{equation}
with $\mu=\nu$ we have $\Psi^{\mu\mu}=0$, and the used wave function
is as follows
\begin{equation}
\Psi^{\mu\nu}=\left(\begin{array}{cccc}
0 & \Psi^{01} & \Psi^{02} & \Psi^{03}\\
-\Psi^{01} & 0 & \Psi^{12} & \Psi^{13}\\
-\Psi^{02} & -\Psi^{12} & 0 & \Psi^{23}\\
-\Psi^{03} & -\Psi^{13} & -\Psi^{23} & 0
\end{array}\right).\label{eq:2}
\end{equation}

With $F_{\mu\nu}=\partial_{\mu}A_{\nu}-\partial_{\nu}A_{\mu}$, $F_{\mu\nu}$
being the electromagnetic field tensor, defining the quadri-vector
of the electromagnetic field by $A_{\mu}=(A_{0},\overrightarrow{A})$.
In presence of an external electromagnetic field, the wave function
in commutative space-time obeys the following covariant equation \cite{key-23}
\begin{equation}
\left(\Gamma\pi+mB\right)_{\;\rho\sigma}^{\mu\nu}\Psi^{\rho\sigma}=0,\label{eq:3}
\end{equation}
with $\pi_{\lambda}=p_{\lambda}+eA_{\lambda}$, $p_{\lambda}=i\left(\partial/\partial x^{\lambda}\right)$
and the matrices $\left(\Gamma^{\lambda}\right)_{\;\rho\sigma}^{\mu\nu}$
and $B_{\;\rho\sigma}^{\mu\nu}$ are given by
\begin{equation}
B_{\;\rho\sigma}^{\mu\nu}=g_{\;\rho}^{\mu}g_{\;\sigma}^{\nu}\text{,}\label{eq:6-1}
\end{equation}
\begin{equation}
\left(\Gamma^{\lambda}\right)_{\;\rho\sigma}^{\mu\nu}=-\frac{4}{3}\gamma^{\lambda}g_{\;\rho}^{\mu}g_{\;\sigma}^{\nu}-\frac{1}{3}\gamma^{\lambda}\left(\gamma^{\mu}\gamma_{\sigma}g_{\;\rho}^{\nu}-\gamma^{\nu}\gamma_{\sigma}g_{\;\rho}^{\mu}\right)+\frac{1}{3}\left(\gamma^{\mu}g_{\;\rho}^{\lambda}g_{\;\sigma}^{\nu}-\gamma^{\nu}g_{\;\rho}^{\lambda}g_{\;\sigma}^{\mu}-g^{\lambda\mu}\gamma_{\rho}g_{\;\sigma}^{\nu}+g^{\lambda\nu}\gamma_{\rho}g_{\;\sigma}^{\mu}\right).\label{eq:5}
\end{equation}

The gamma matrices obey the following Clifford covariant algebra
\begin{equation}
\begin{array}{cc}
\gamma^{\mu}\gamma^{\nu}+\gamma^{\nu}\gamma^{\mu}=2g^{\mu\nu}, & \sigma^{\mu\nu}=\frac{1}{2}i\left(\gamma^{\mu}\gamma^{\nu}-\gamma^{\nu}\gamma^{\mu}\right)\end{array},\label{eq:7}
\end{equation}
where $\gamma^{0}=\left(\begin{array}{cc}
1 & 0\\
0 & -1
\end{array}\right),\:\gamma^{5}=\gamma^{0}\gamma^{1}\gamma^{2}\gamma^{3},\quad\overrightarrow{\gamma}=\left(\begin{array}{cc}
0 & \overrightarrow{\sigma}\\
-\overrightarrow{\sigma} & 0
\end{array}\right)$. Taking into account that $\gamma^{0}$ is Hermitian, and $\overrightarrow{\gamma}$
is anti-Hermitian. 

Using Eqs.(\ref{eq:6-1}, \ref{eq:5}), the Fisk\textendash Tait equation
in a more detailed form is given by
\begin{equation}
\begin{array}{c}
m\Psi^{\mu\nu}-\frac{4}{3}\gamma^{\lambda}\pi_{\lambda}\Psi^{\mu\nu}+\frac{1}{3}\gamma^{\lambda}\pi_{\lambda}\left(\gamma^{\nu}\gamma_{\sigma}\Psi^{\mu\sigma}-\gamma^{\mu}\gamma_{\sigma}\Psi^{\nu\sigma}\right)+\frac{1}{3}\left(\gamma^{\mu}\pi_{\rho}\Psi^{\rho\nu}-\gamma^{\nu}\pi_{\rho}\Psi^{\rho\mu}-\gamma_{\rho}\pi^{\mu}\Psi^{\rho\nu}+\gamma_{\rho}\pi^{\nu}\Psi^{\rho\mu}\right)=0\end{array}.\label{eq:11}
\end{equation}

The employed wave function tensor-spinor is decomposed in two wave
vectors components as
\begin{equation}
\Psi^{\mu\nu}=\left(\begin{array}{c}
\overrightarrow{\psi}\\
\overrightarrow{\chi}
\end{array}\right),\;\begin{array}{cc}
\begin{array}{c}
\overrightarrow{\psi}=(\Psi^{01},\Psi^{02},\Psi^{03}),\end{array} & \overrightarrow{\chi}=(\Psi^{23},\Psi^{31},\Psi^{12})\end{array}.\label{eq:12}
\end{equation}

Knowing that the spin-3/2 parts of the wave function are
\begin{equation}
\overrightarrow{\psi}_{\:3/2}=\overrightarrow{\psi}+\frac{1}{3}\overrightarrow{\gamma}(\overrightarrow{\gamma}\overrightarrow{\psi}),\;\overrightarrow{\chi}_{\:3/2}=\overrightarrow{\chi}+\frac{1}{3}\overrightarrow{\gamma}(\overrightarrow{\gamma}\overrightarrow{\chi}),\label{eq:13}
\end{equation}
with the subsidiary conditions $\overrightarrow{\gamma}\overrightarrow{\psi}_{3/2}=\overrightarrow{\gamma}\overrightarrow{\chi}_{3/2}=0$.

In case of a constant real magnetic field, we have the following expression
\begin{equation}
F_{\mu\nu}=\left(\begin{array}{cccc}
0 & 0 & 0 & 0\\
0 & 0 & \mathcal{\mathbb{\mathcal{H}}} & 0\\
0 & -\mathcal{H} & 0 & 0\\
0 & 0 & 0 & 0
\end{array}\right),\label{eq:8}
\end{equation}
where $\mathcal{H}$ is a constant real magnetic field oriented along
the axis (Oz). This present physical system is often referred to as
the Landau system. For this problem, convenient choice of the electromagnetic
field vector is the following symmetric gauge
\begin{equation}
\overrightarrow{A}=\frac{\mathcal{H}}{2}\left(-y,x,0\right),\;\text{with }\:A_{0}=0.\label{eq:10}
\end{equation}

For simplicity, we may choose Landau gauge to reduce $\overrightarrow{A}$
to one component such that either
\begin{equation}
\overrightarrow{A}=\mathcal{\mathbb{\mathcal{H}}}\left(0,x,0\right),\:\text{or}\:\overrightarrow{A}=\mathcal{\mathbb{\mathcal{H}}}\left(-y,0,0\right).\label{eq:11-1-1}
\end{equation}

It should be stressed out that the expression of $\overrightarrow{A}$
in the symmetric and Landau gauges is obtained by using Eq.(\ref{eq:8}).
Therefore, the Fisk\textendash Tait equation (\ref{eq:3}) goes to
\begin{equation}
\left(\Gamma^{\lambda}\right)_{\;\rho\sigma}^{\mu\nu}p_{\lambda}\Psi^{\rho\sigma}-eA_{k}\left(\Gamma^{k}\right)_{\;\rho\sigma}^{\mu\nu}\Psi^{\rho\sigma}+mB_{\;\rho\sigma}^{\mu\nu}\Psi^{\rho\sigma}=0,\;k=1,2,3,\label{eq:4-1-1}
\end{equation}
with
\begin{equation}
\left(\Gamma^{k}\right)_{\;\rho\sigma}^{\mu\nu}=-\frac{4}{3}\gamma^{k}g_{\;\rho}^{\mu}g_{\;\sigma}^{\nu}-\frac{1}{3}\gamma^{k}\left(\gamma^{\mu}\gamma_{\sigma}g_{\;\rho}^{\nu}-\gamma^{\nu}\gamma_{\sigma}g_{\;\rho}^{\mu}\right)+\frac{1}{3}\left(\gamma^{\mu}g_{\;\rho}^{k}g_{\;\sigma}^{\nu}-\gamma^{\nu}g_{\;\rho}^{k}g_{\;\sigma}^{\mu}-g^{k\mu}\gamma_{\rho}g_{\;\sigma}^{\nu}+g^{k\nu}\gamma_{\rho}g_{\;\sigma}^{\mu}\right).\label{eq:20-1-1}
\end{equation}

In the general case of a non-constant magnetic field, we introduce
a function depending on $x$ in the Landau gauge as $A_{2}=x\mathcal{\mathbb{\mathcal{H}}}f(x)$
which gives us a non-constant magnetic field. The magnetic field can
be calculated easily using $\overrightarrow{\mathcal{\mathbb{\mathcal{H}}}}=\overrightarrow{\nabla}\times\overrightarrow{A}$
which gives \cite{key-24}
\begin{equation}
\overrightarrow{\mathcal{\mathbb{\mathcal{H}}}}(x)=\left(x\mathcal{\mathbb{\mathcal{H}}}\frac{d}{dx}f\left(x\right)+\mathcal{\mathbb{\mathcal{H}}}f\left(x\right)\right)\overrightarrow{e}_{3}.\label{eq:12-1-1}
\end{equation}

The above equation represents a special kind of non-constant magnetic
fields. If we specify $f(x)$ we obtain different classes of the non-constant
magnetic field. If take $f(x)=1$ in this case we get a constant magnetic
field. Let\textquoteright s consider $f(x)=\frac{1}{x}\left(1-e^{-x}\right)$,
to obtain an exponentially decaying magnetic field. Of course, a multitude
of other possibilities exist, so that Eq.(\ref{eq:12-1-1}) goes to
\begin{equation}
\overrightarrow{\mathcal{\mathbb{\mathcal{H}}}}(x)=\left(\mathcal{\mathbb{\mathcal{H}}}e^{-x}\right)\overrightarrow{e}_{3}.\label{eq:13-2}
\end{equation}

\subsection{{\normalsize{}THE FISK}\textendash {\normalsize{}TAIT EQUATION IN
NONCOMMUTATIVE SPACE-TIME}}

The noncommutative geometry \cite{key-25,key-26,key-27} is the theory
in which space may not commute anymore, so that in a 4d noncommutative
space-time (NCST). The usual space-time coordinates $x^{\mu}$, are
replaced by the operators $\hat{x}^{\mu}$, which satisfy the commutation
relations \cite{key-28} 
\begin{equation}
\left[\hat{x}^{\mu},\hat{x}^{\nu}\right]=i\Theta^{\mu\nu},\quad(\mu\nu=0,...,3),\label{eq:14}
\end{equation}
where the noncommutativity parameter is a real-valued antisymmetric
constant matrix with a dimension of $length{}^{2}$. And the uncertainty
relations $\triangle\hat{x}^{\mu}\triangle\hat{x}^{\nu}\geq\left|\Theta^{\mu\nu}\right|/2$,
are compatible with Eq.(\ref{eq:14}). For some studies about noncommutative
systems concerning the noncommutative parameter, there are some experiments
that have linked the noncommutative parameter scale to $\Theta\approx4.10^{-40}m^{2}$(which
corresponds to $\Theta\approx10^{-2}Tev^{-2}$). This value corresponds
to the upper bound on the fundamental coordinate scale \cite{key-29},
this bound will be automatically canceled when the magnetic field
used in the experiment is weak $B\approx5mG$. For our needs, in the
context of the field theory, it is more practical to introduce the
noncommutative scale $\varLambda_{nc}$ by \cite{key-30}
\begin{equation}
\Theta^{\mu\nu}=\frac{c^{\mu\nu}}{\varLambda_{nc}},\label{eq:15}
\end{equation}
with $c^{\mu\nu}$ is a dimensionless antisymmetric tensor, where
the components are $\mathcal{O}(1)$. $\varLambda_{nc}$ represents
a characteristic energy scale for the noncommutative theory which
is necessarily quite large.

We would like to note that, it would be quite well to approach the
problem via the Seiberg\textendash Witten maps if we have an electromagnetic
field interaction. But while we have a magnetic field interaction,
we will follow a quite standard approach that has been widely used
in the literature on noncommutative quantum mechanics (NCQM); which
depends on obtaining a noncommutative version of a given field theory,
and based on replacing the product of the fields by the Moyal-Weyl
product ($\star$\textendash product) defined as{\small{} \cite{key-31,key-32,key-33}
\begin{equation}
(f\star g)(x)=\exp[\frac{i}{2}\Theta_{\mu\nu}\partial_{x_{\mu}}\partial_{x_{\nu}}]f(x_{\mu})g(x_{\nu})|_{x_{\mu}=x_{\nu}}=f(x)g(x)+\sum_{n=1}\left(\frac{1}{n!}\right)\left(\frac{i}{2}\right)^{n}\Theta^{\mu_{1}\nu_{1}}...\Theta^{\mu_{n}\nu_{n}}\partial_{\mu_{1}}...\partial_{\mu_{k}}f(x)\partial_{\nu_{1}}...\partial_{\nu_{k}}g(x).\label{eq:16}
\end{equation}
}{\small \par}

We want to emphasize that the non-commutative field theories for the
low energies ($\Theta E^{2}<1$) or the slowly varying fields reduce
to their commutative version ($\Theta=0$) due to the nature of the
Moyal $\star$\textendash product.

It is well known, that the noncommutative coordinates operator $\hat{x}^{\mu}$
are linked to the commutative one through the Heisenberg\textendash Weyl
algebra in terms of the following Darboux transformation
\begin{equation}
\hat{x}^{\mu}=x^{\mu}-\frac{\Theta^{\mu\nu}}{2}p_{\nu},\;\hat{p}^{\mu}=p^{\mu}.\label{eq:18-1}
\end{equation}

The variables $x^{\mu}$, $p^{\mu}$ satisfy the usual canonical commutation
relations in QM. In what follows, we study the dynamical equation
in NCST for the case of constant and non-constant magnetic fields
using the Landau gauge. The Fisk\textendash Tait equation in NCST
is written as follows 
\begin{equation}
\left\{ (\Gamma^{\lambda})_{\;\rho\sigma}^{\mu\nu}p_{\lambda}+mB_{\;\rho\sigma}^{\mu\nu}\right\} \Psi^{\rho\sigma}+e(\Gamma^{\lambda})_{\;\rho\sigma}^{\mu\nu}A_{\lambda}(x)\star\Psi^{\rho\sigma}\left(x\right)=0,\label{eq:17}
\end{equation}
where the $\star$\textendash product is a realization of algebra
(\ref{eq:16}). One can write the noncommutative part as

{\small{}
\begin{equation}
\begin{array}{c}
\left(\Gamma^{\lambda}\right)_{\;\rho\sigma}^{\mu\nu}A_{\lambda}(x)\star\Psi^{\rho\sigma}\left(x\right)\thickapprox\left(\Gamma^{\lambda}\right)_{\;\rho\sigma}^{\mu\nu}A_{\lambda}\Psi^{\rho\sigma}+\frac{i}{2}\Theta^{\alpha\beta}\left(\Gamma^{\lambda}\right)_{\;\rho\sigma}^{\mu\nu}\partial_{\alpha}\left\{ A_{\lambda}\right\} \partial_{\beta}\Psi^{\rho\sigma}-\frac{1}{8}\left(\Theta^{\alpha\beta}\right)^{2}\left(\partial_{\alpha}\right)^{2}\left\{ (\Gamma^{\lambda})_{\;\rho\sigma}^{\mu\nu}A_{\lambda}\right\} \left(\partial_{\beta}\right)^{2}\Psi^{\rho\sigma}\\
+...\underset{n=3}{\sum}\left(\frac{1}{n!}\right)\left(\frac{i}{2}\right)^{n}\Theta^{a_{1}b_{1}}...\Theta^{a_{n}b_{n}}\partial_{a_{1}}...\partial_{a_{k}}\left\{ (\Gamma^{\lambda})_{\;\rho\sigma}^{\mu\nu}A_{\lambda}\right\} \partial_{b_{1}}...\partial_{b_{k}}\Psi^{\rho\sigma}.
\end{array}\label{eq:18}
\end{equation}
}{\small \par}

However, one must pay attention to the ordering issues that can arise.
In addition, with the help of Eq.(\ref{eq:18-1}), one can check that
the expression (\ref{eq:18}) can be rewritten as
\begin{equation}
\left(\Gamma^{\lambda}\right)_{\;\rho\sigma}^{\mu\nu}A_{\lambda}(x)\star\Psi^{\rho\sigma}\left(x\right)=\left(\Gamma^{\lambda}\right)_{\;\rho\sigma}^{\mu\nu}A_{\lambda}(\hat{x}^{\mu})\Psi^{\rho\sigma}\left(\hat{x}\right).\label{eq:22-3}
\end{equation}

\subsubsection{\texttt{\textup{CASE OF A CONSTANT MAGNETIC FIELD}}}

To first order of $\Theta$, taking into account the case of potential
$A_{2}=\mathcal{\mathbb{\mathcal{H}}}x$, the derivations in algebra
(\ref{eq:18}) roughly stop in the first order of $\Theta$, then
the approximation gives
\begin{equation}
\left\{ (\Gamma^{\lambda})_{\;\rho\sigma}^{\mu\nu}p_{\lambda}+mB_{\;\rho\sigma}^{\mu\nu}\right\} \Psi^{\rho\sigma}-e\mathcal{\mathbb{\mathcal{H}}}x\left(\Gamma^{2}\right)_{\;\rho\sigma}^{\mu\nu}\Psi^{\rho\sigma}-\frac{ie}{2}\mathcal{\mathbb{\mathcal{H}}}\Theta^{\alpha\beta}\left(\Gamma^{2}\right)_{\;\rho\sigma}^{\mu\nu}\partial_{\alpha}\left\{ x\right\} \partial_{\beta}\Psi^{\rho\sigma}+0(\Theta^{2})=0.\label{eq:19}
\end{equation}

We can see to this order, our noncommutative Fisk\textendash Tait
equation suggests the following quantity as a noncommutative correction
\begin{equation}
\frac{ie}{2}\mathcal{\mathbb{\mathcal{H}}}\Theta^{\alpha\beta}\left(\Gamma^{2}\right)_{\;\rho\sigma}^{\mu\nu}\left\{ x\right\} \partial_{\beta}\Psi^{\rho\sigma}=\frac{ie}{2}\mathcal{\mathbb{\mathcal{H}}}\Theta^{1\beta}\left(\Gamma^{2}\right)_{\;\rho\sigma}^{\mu\nu}\partial_{\beta}\Psi^{\rho\sigma},\label{eq:20}
\end{equation}
the most important feature of this approximation is its simple character,
which allows one to work simply to test the noncommutativity effect
on the total charge through the continuity equation later, without
having to worry about the issues coming from the arising series of
the $\star$\textendash product.

\subsubsection{\texttt{\textup{CASE OF A NON-CONSTANT MAGNETIC FIELD}}}

Using Eq.(\ref{eq:11-1-1}) and Eq.(\ref{eq:13-2}) would lead to
a non-constant magnetic field $A_{2}=\mathcal{\mathbb{\mathcal{H}}}xe^{-x}$,
which provides a noncommutative Fisk\textendash Tait equation for
all orders of $\Theta$, as follows
\begin{equation}
\left\{ (\Gamma^{\lambda})_{\;\rho\sigma}^{\mu\nu}\pi_{\lambda}+mB_{\;\rho\sigma}^{\mu\nu}\right\} \Psi^{\rho\sigma}-e(\Gamma^{2})_{\;\rho\sigma}^{\mu\nu}\underset{n=1}{\sum}\left(\frac{1}{n!}\right)\left(\frac{i}{2}\right)^{n}\Theta^{a_{1}b_{1}}...\Theta^{a_{n}b_{n}}\partial_{a_{1}}...\partial_{a_{k}}\left\{ \mathcal{\mathbb{\mathcal{H}}}xe^{-x}\right\} \partial_{b_{1}}...\partial_{b_{k}}\Psi^{\rho\sigma}=0,\label{eq:21}
\end{equation}
below we discuss some cases of the noncommutative correction part:

To first order of $\Theta$ ($n=1$), the approximation yields 
\begin{equation}
\mathcal{C}_{1}=\frac{i}{2}\Theta^{\alpha\beta}\left(\Gamma^{2}\right)_{\;\rho\sigma}^{\mu\nu}\partial_{\alpha}\left\{ \mathcal{\mathbb{\mathcal{H}}}xe^{-x}\right\} \partial_{\beta}\Psi^{\rho\sigma}=\frac{i}{2}\mathcal{\mathbb{\mathcal{H}}}\Theta^{1\beta}\left(\Gamma^{2}\right)_{\;\rho\sigma}^{\mu\nu}\left(1-x\right)e^{-x}\partial_{\beta}\Psi^{\rho\sigma}.\label{eq:25-a}
\end{equation}

To second order of $\Theta$ ($n=2$), with $a_{1}=a_{2}=1$, $b_{1}=b_{2}=\beta$,
the approximation yields
\begin{equation}
\mathcal{C}_{2}=\mathcal{C}_{1}+\frac{1}{8}\mathcal{\mathbb{\mathcal{H}}}\left(\Theta^{1\beta}\right)^{2}(\Gamma^{\lambda})_{\;\rho\sigma}^{\mu\nu}\left(2-x\right)e^{-x}\left(\partial_{\beta}\right)^{2}\Psi^{\rho\sigma}.\label{eq:25-b}
\end{equation}

To third order of $\Theta$ ($n=3$), with $a_{1}=a_{2}=a_{3}=1$,
$b_{1}=b_{2}=b_{3}=\beta$, the approximation yields
\begin{equation}
\mathcal{C}_{3}=\mathcal{C}_{2}+\frac{i}{48}\mathcal{\mathbb{\mathcal{H}}}\left(\Theta^{1\beta}\right)^{3}(\Gamma^{\lambda})_{\;\rho\sigma}^{\mu\nu}\left(3-x\right)e^{-x}\left(\partial_{\beta}\right)^{3}\Psi^{\rho\sigma}.\label{eq:25-c}
\end{equation}

To calculate the $n{}^{th}$ order term, we consider that, $a_{1}=a_{2}=...=a_{n}=1$,
$b_{1}=b_{2}=...=b_{n}=\beta$, we have
\begin{equation}
\mathcal{C}_{n}=\mathcal{C}_{n-1}-\frac{1}{n!}\left(\frac{i}{2}\right)^{n}\left(\Theta^{1\beta}\right)^{n}(\Gamma^{\lambda})_{\;\rho\sigma}^{\mu\nu}\left(n-x\right)e^{-x}\left(\partial_{\beta}\right)^{n}\Psi^{\rho\sigma}.\label{eq:25-d}
\end{equation}

From the comparison between commutative and noncommutative systems,
it is widely obvious that the $\star$\textendash product causes noncommutative
corrections on the Fisk\textendash Tait equation due to the presence
of the magnetic field. This correction is controlled by the form and
type of the magnetic field. It can be shown that, in NCST, if we take
a constant magnetic field, the order of the noncommutativity cannot
be higher than the first order. On the other hand, when we take a
non-constant magnetic field (at least to the form of the non-constant
magnetic field we have considered), It causes high orders in noncommutativity,
which produces infinite series of $\Theta$.

In the noncommutative Fisk\textendash Tait equation, the appearance
of terms proportional to explicit $\Theta$ parameter, in fact, because
of the noncommutativity effect on the dynamical equation. In the case
of a constant magnetic field, we consider the appeared correction
term as a kind of potential. Therefore, we will see if it remains
the total charge obtained from the probability density indefinite
or no.

It should also be noted that the case of higher dimensions of $\Theta$
and a more general algebra are complicated, all along with our physical
systems involving both electrical and magnetic potentials.

\section{the continuity equation in two types of space-time }

\subsection{{\normalsize{}THE CONTINUITY EQUATION IN COMMUTATIVE CASE}}

The Fisk\textendash Tait equation in the commutative space-time is
given by Eq.(\ref{eq:11}), with the following subsidiary conditions
\begin{equation}
\gamma_{\mu}\gamma_{\nu}\Psi^{\mu\nu}=0\text{,}\:\;\pi_{\mu}\gamma_{\nu}\Psi^{\mu\nu}=0.\label{eq:23}
\end{equation}

The complex conjugate of Eq.(\ref{eq:11}) is

\begin{equation}
\begin{array}{c}
m\Psi^{\dagger\mu\nu}-\frac{4}{3}\Psi^{\dagger\mu\nu}\pi_{\lambda}^{\ast}\gamma^{\lambda\dagger}+\frac{1}{3}\left(-\Psi^{\dagger\nu\sigma}\gamma_{\sigma}^{\dagger}\gamma^{\dagger\mu}\gamma^{\dagger\lambda}+\Psi^{\dagger\mu\sigma}\gamma_{\sigma}^{\dagger}\gamma^{\dagger\nu}\gamma^{\dagger\lambda}\right)\pi_{\lambda}^{\ast}\\
+\frac{1}{3}\left(\Psi^{\dagger\rho\nu}\pi_{\rho}^{\ast}\gamma^{\dagger\mu}-\Psi^{\dagger\rho\mu}\pi_{\rho}^{\ast}\gamma^{\dagger\nu}-\Psi^{\dagger\rho\nu}\pi^{\ast\mu}\gamma_{\rho}^{\dagger}+\Psi^{\dagger\rho\mu}\pi^{\ast\nu}\gamma_{\rho}^{\dagger}\right)=0.
\end{array}\label{eq:24}
\end{equation}

Here $\ast$ and $\dagger$ stand for the complex conjugation of $\pi$
(even the potentials) and the wave functions, tensors respectively.

Of multiplying $\Psi_{\mu\nu}^{\dagger}$ by Eq.(\ref{eq:11}), we
have 
\begin{equation}
\begin{array}{c}
m\Psi_{\mu\nu}^{\dagger}\Psi^{\mu\nu}-\frac{4}{3}\Psi_{\mu\nu}^{\dagger}\gamma^{\lambda}\pi_{\lambda}\Psi^{\mu\nu}+\frac{1}{3}\Psi_{\mu\nu}^{\dagger}\gamma^{\lambda}\pi_{\lambda}\left(-\gamma^{\mu}\gamma_{\sigma}\Psi^{\nu\sigma}+\gamma^{\nu}\gamma_{\sigma}\Psi^{\mu\sigma}\right)\\
+\frac{1}{3}\left(\Psi_{\mu\nu}^{\dagger}\gamma^{\mu}\pi_{\rho}\Psi^{\rho\nu}-\Psi_{\mu\nu}^{\dagger}\gamma^{\nu}\pi_{\rho}\Psi^{\rho\mu}-\Psi_{\mu\nu}^{\dagger}\gamma_{\rho}\pi^{\mu}\Psi^{\rho\nu}+\Psi_{\mu\nu}^{\dagger}\gamma_{\rho}\pi^{\nu}\Psi^{\rho\mu}\right)=0.
\end{array}\label{eq:25}
\end{equation}

Multiplying Eq.(\ref{eq:24}) by $\Psi_{\mu\nu}$, we obtain
\begin{equation}
\begin{array}{c}
m\Psi^{\dagger\mu\nu}\Psi_{\mu\nu}-\frac{4}{3}\Psi^{\dagger\mu\nu}\pi_{\lambda}^{\ast}\gamma^{\lambda\dagger}\Psi_{\mu\nu}+\frac{1}{3}\left(-\Psi^{\dagger\nu\sigma}\gamma_{\sigma}^{\dagger}\gamma^{\dagger\mu}\gamma^{\dagger\lambda}+\Psi^{\dagger\mu\sigma}\gamma_{\sigma}^{\dagger}\gamma^{\dagger\nu}\gamma^{\dagger\lambda}\right)\pi_{\lambda}^{\ast}\Psi_{\mu\nu}\\
+\frac{1}{3}\left(\Psi^{\dagger\rho\nu}\pi_{\rho}^{\ast}\gamma^{\dagger\mu}\Psi_{\mu\nu}-\Psi^{\dagger\rho\mu}\pi_{\rho}^{\ast}\gamma^{\dagger\nu}\Psi_{\mu\nu}-\Psi^{\dagger\rho\nu}\pi^{\ast\mu}\gamma_{\rho}^{\dagger}\Psi_{\mu\nu}+\Psi^{\dagger\rho\mu}\pi^{\ast\nu}\gamma_{\rho}^{\dagger}\Psi_{\mu\nu}\right)=0.
\end{array}\label{eq:26}
\end{equation}

With some minor simplifications and according to the subtraction of
Eq.(\ref{eq:25}) from Eq.(\ref{eq:26}), and by taking $\pi_{\lambda}=i\partial_{\lambda}$
for simplicity, which means considering the Fisk\textendash Tait equation
without the electromagnetic interaction, we find
\begin{equation}
\partial_{\lambda}\mathcal{J}^{\lambda}=\partial_{\lambda}\left(-\frac{4}{3}\Psi^{\dagger\mu\nu}\gamma^{\lambda}\Psi^{\mu\nu}-\frac{2}{3}\Psi^{\dagger\mu\nu}\gamma^{\lambda}\gamma^{\mu}\gamma_{\sigma}\Psi^{\nu\sigma}+\frac{2}{3}\Psi^{\dagger\mu\nu}\gamma^{\mu}\Psi^{\lambda\nu}-\frac{2}{3}\Psi^{\dagger\lambda\nu}\gamma_{\rho}\Psi^{\rho\nu}\right)=0,\label{eq:28}
\end{equation}
the above equation is the continuity equation, where the current density
quadri-vector $\mathcal{J}^{\lambda}=(\mathcal{J}^{0},\mathcal{J}^{i})$
is
\begin{equation}
\mathcal{J}^{\lambda}=-\frac{2}{3}\left(2\overline{\Psi}^{\mu\nu}\gamma^{\lambda}\Psi_{\mu\nu}+\overline{\Psi}^{\mu\nu}\gamma^{\lambda}\gamma_{\mu}\gamma^{\sigma}\Psi_{\nu\sigma}-\overline{\Psi}^{\mu\nu}\gamma_{\mu}\Psi_{\nu}^{\lambda}+\overline{\Psi}^{\lambda\nu}\gamma^{\rho}\Psi_{\rho\nu}\right),\qquad\overline{\Psi}^{\mu\nu}=\Psi^{\dagger\mu\nu}\gamma^{0}.\label{eq:29}
\end{equation}

The corresponding conserved quantity is the total probability 

\begin{equation}
\mathcal{J}^{0}=-\frac{2}{3}\left(2\Psi^{\dagger\mu\nu}\Psi_{\mu\nu}+\Psi^{\dagger\mu\nu}\gamma_{\mu}\gamma^{\sigma}\Psi_{\nu\sigma}-\Psi^{\dagger\mu\nu}\gamma^{0}\gamma_{\mu}\Psi_{\nu}^{0}+\Psi^{\dagger0\nu}\gamma^{0}\gamma^{\rho}\Psi_{\rho\nu}\right),\label{eq:29-1}
\end{equation}
where $\mathcal{J}^{0}$ is the probability density for finding a
particle at a particular position, and $\mathcal{J}^{i}$ is the current
density of the spin-3/2 particle. It is clear that the total charge
$\mathcal{Q}$ obtained from the probability density is indefinite
\begin{equation}
\mathcal{Q}=\int\mathcal{J}^{0}d^{3}x=\int d^{3}x\left(-\frac{4}{3}\Psi^{\dagger ij}\Psi^{ij}+2\Psi^{\dagger0i}\Psi^{0i}-\frac{2}{3}\Psi^{\dagger ij}\gamma_{i}\gamma^{k}\Psi_{jk}+\frac{2}{3}\Psi^{\dagger oi}\gamma_{i}\gamma^{j}\Psi_{ki}\right),\label{eq:38-1}
\end{equation}
using Eqs.(\ref{eq:12}, \ref{eq:13}), we obtain
\begin{equation}
\mathcal{Q}=2\int d^{3}x\left(\overrightarrow{\psi}_{\:3/2}^{\dagger}\overrightarrow{\psi}_{\:3/2}-\overrightarrow{\chi}_{\:3/2}^{\dagger}\overrightarrow{\chi}_{\:3/2}\right).\label{eq:38-2}
\end{equation}

For all the spin 3/2 equations of motion, the problem of total charge
indefiniteness remained \cite{key-12}, knowing that, in the case
of Gupta equation \cite{key-14} with some conditions, the charge
density is positive-definite. However, generally the total charge
in the free theory is indefinite which means, that the condition for
the positivity of charge is first noted, then it is shown that this
condition has to be violated if causality of propagation is to be
ensured. The lack of a positive definite probability density suggests
that many of the problems that plague attempts to quantize spin-3/2
fields are also to be found in the classical field equations.

It should be noted, once considering the Fisk\textendash Tait equation
in interaction with EMF (which means taking $\pi_{\lambda}=p_{\lambda}+eA_{\lambda}$)
in the calculations of the continuity equation, this, in turn, leads
to getting a correction term (of probability density type) in the
continuity equation and the total charge expressions. We are going
to show this in the next subsection together with the noncommutativity
influence.

\subsection{{\normalsize{}THE CONTINUITY EQUATION IN NONCOMMUTATIVE CASE}}

The Fisk\textendash Tait equation in NCPS is given by Eq.(\ref{eq:19}),
and its complex conjugate is given by
\begin{equation}
\begin{array}{c}
m\Psi^{\dagger\mu\nu}-\frac{4}{3}\Psi^{\dagger\mu\nu}p_{\lambda}^{\ast}\gamma^{\lambda\dagger}+\frac{1}{3}\left(-\Psi^{\dagger\nu\sigma}\gamma_{\sigma}^{\dagger}\gamma^{\dagger\mu}\gamma^{\dagger\lambda}+\Psi^{\dagger\mu\sigma}\gamma_{\sigma}^{\dagger}\gamma^{\dagger\nu}\gamma^{\dagger\lambda}\right)p_{\lambda}^{\ast}\\
+\frac{1}{3}\left(\Psi^{\dagger\rho\nu}p_{\rho}^{\ast}\gamma^{\dagger\mu}-\Psi^{\dagger\rho\mu}p_{\rho}^{\ast}\gamma^{\dagger\nu}-\Psi^{\dagger\rho\nu}p^{\ast\mu}\gamma_{\rho}^{\dagger}+\Psi^{\dagger\rho\mu}p^{\ast\nu}\gamma_{\rho}^{\dagger}\right)\\
-e\mathcal{\mathbb{\mathcal{H}}}x\Psi^{\dagger\rho\sigma}\left(\Gamma^{\dagger2}\right)_{\;\rho\sigma}^{\mu\nu}+\frac{ie}{2}\mathcal{\mathbb{\mathcal{H}}}\Theta^{1\beta}\left(\partial_{\beta}\Psi^{\rho\sigma}\right)^{\dagger}\left(\Gamma^{2\dagger}\right)_{\;\rho\sigma}^{\mu\nu}=0.
\end{array}\label{eq:31}
\end{equation}

Noting that it is enough to consider the simplest case concerning
the magnetic field $A_{2}=\mathcal{\mathbb{\mathcal{H}}}x$, to study
the total charge within the noncommutativity at an explicitly solvable
model. With $p_{\lambda}=i\partial_{\lambda}$, and multiplying $\Psi_{\mu\nu}^{\dagger}$
by Eq.(\ref{eq:19}) and Eq.(\ref{eq:31}) by $\Psi_{\mu\nu}$. Therefore,
by subtracting $\Psi_{\mu\nu}^{\dagger}$Eq.(\ref{eq:19}) from Eq.(\ref{eq:31})$\Psi_{\mu\nu}$,
we obtain
\begin{equation}
\begin{array}{c}
i\partial_{\lambda}\left[-\frac{4}{3}\overline{\Psi}^{\mu\nu}\gamma^{\lambda}\Psi_{\mu\nu}-\frac{2}{3}\overline{\Psi}^{\mu\nu}\gamma^{\lambda}\gamma_{\mu}\gamma_{\sigma}\Psi_{\nu\sigma}+\frac{2}{3}\overline{\Psi}^{\mu\nu}\gamma_{\mu}\Psi_{\nu}^{\lambda}-\frac{2}{3}\overline{\Psi}^{\lambda\nu}\gamma^{\rho}\Psi_{\rho\nu}\right]\\
+e\mathcal{\mathbb{\mathcal{H}}}x\left[\Psi^{\dagger\rho\sigma}\left(\Gamma^{\dagger2}\right)_{\;\rho\sigma}^{\mu\nu}\Psi_{\mu\nu}-\Psi_{\mu\nu}^{\dagger}\left(\Gamma^{2}\right)_{\;\rho\sigma}^{\mu\nu}\Psi^{\rho\sigma}\right]\\
-\frac{ie}{2}\mathcal{\mathbb{\mathcal{H}}}\Theta^{1\beta}\left[\Psi_{\mu\nu}^{\dagger}\left(\Gamma^{2}\right)_{\;\rho\sigma}^{\mu\nu}\partial_{\beta}\Psi^{\rho\sigma}+\Psi^{\dagger\rho\sigma}\partial_{\beta}\left(\Gamma^{\dagger2}\right)_{\;\rho\sigma}^{\mu\nu}\Psi_{\mu\nu}\right]=0,
\end{array}\label{eq:32}
\end{equation}
contracting the above equation as follows
\begin{equation}
\partial_{\lambda}\mathcal{J}^{\lambda}+\zeta^{2}+\zeta_{nc}^{2}=0,\label{eq:42-11}
\end{equation}
the Eq.(\ref{eq:42-11}) will be recognized as the noncommutative
continuity equation, denoting the separate terms in it as follows
\begin{equation}
\begin{array}{ccc}
\mathcal{J}^{\lambda} & = & -\frac{4}{3}\overline{\Psi}^{\mu\nu}\gamma^{\lambda}\Psi_{\mu\nu}-\frac{2}{3}\overline{\Psi}^{\mu\nu}\gamma^{\lambda}\gamma_{\mu}\gamma_{\sigma}\Psi_{\nu\sigma}+\frac{2}{3}\overline{\Psi}^{\mu\nu}\gamma_{\mu}\Psi_{\nu}^{\lambda}-\frac{2}{3}\overline{\Psi}^{\lambda\nu}\gamma^{\rho}\Psi_{\rho\nu}\\
\zeta^{2} & = & e\mathcal{\mathbb{\mathcal{H}}}x\left[\Psi^{\dagger\rho\sigma}\left(\Gamma^{\dagger2}\right)_{\;\rho\sigma}^{\mu\nu}\Psi_{\mu\nu}-\Psi_{\mu\nu}^{\dagger}\left(\Gamma^{2}\right)_{\;\rho\sigma}^{\mu\nu}\Psi^{\rho\sigma}\right]\qquad\qquad\qquad\\
\zeta_{nc}^{2} & = & \frac{-e}{2}\mathcal{\mathbb{\mathcal{H}}}\Theta^{1\beta}\left[\Psi_{\mu\nu}^{\dagger}\left(\Gamma^{2}\right)_{\;\rho\sigma}^{\mu\nu}\partial_{\beta}\Psi^{\rho\sigma}+\Psi^{\dagger\rho\sigma}\partial_{\beta}\left(\Gamma^{\dagger2}\right)_{\;\rho\sigma}^{\mu\nu}\Psi_{\mu\nu}\right]\qquad
\end{array}.\label{eq:33}
\end{equation}

Where $\mathcal{J}^{\lambda}$ is the current density quadri-vector
of the spin-3/2 particle. The quantity $\zeta^{2}$ is a term of the
probability density type; this quantity is related to the effect of
the electromagnetic field interaction on the Fisk\textendash Tait
equation. This quantity emerged merely as a term containing the constant
magnetic field contribution, consequently after extracting the commutative
or noncommutative continuity equation, this contribution being responsible
for generating this probability density quantity. The correction term
$\zeta_{nc}^{2}$ is a new noncommutative quantity of current density
type, the existence of this quantity corresponding to the explicit
space noncommutative parameter which involved in the obtained noncommutative
continuity equation, that is because of the influence of the space
noncommutativity on the spin-3/2 particle motion equation. And such
quantity appeared as a term which is proportional to the noncommutativity
parameter $\Theta$, then after extracting the noncommutative continuity
equation, this appeared term being responsible for generating the
new quantity term of the current density type. Once the magnetic field
is null, the quantities $\zeta^{2}$ and $\zeta_{nc}^{2}$ will disappear.

We see that the total charge $\mathcal{Q}$ obtained from the probability
density $\mathcal{J}^{0}$ still indefinite even when space does not
commute. To demonstrate its indefiniteness, it is enough to consider
the rest frame in which $p=0$. Then Eq.(\ref{eq:23}) reduces to
\begin{equation}
\begin{array}{cc}
\gamma_{i}\gamma_{j}\Psi^{ij}=0\text{, } & \gamma_{i}\Psi^{0i}=0\end{array},\label{eq:44-1}
\end{equation}
so that
\begin{equation}
\mathcal{Q}^{nc}=\mathcal{Q}+\zeta^{2}+\zeta_{nc}^{2}=\int\mathcal{J}^{0}d^{3}x+\zeta^{2}+\zeta_{nc}^{2}.\label{eq:44-2}
\end{equation}

It is easily seen that relations (\ref{eq:44-1}) do not eliminate
the indefiniteness of $\mathcal{Q}^{nc}$.

\section{AN ADDITIONAL CONTRIBUTION}

There are several ideas from the electricity, the magnetism, and nuclear
physics that suggest the contribution of the spin and the current
density. In this section, we try to highlight some of those ideas.
We have $S=(S_{1},S_{2},S_{3})$, knowing that $S_{1},S_{2},S_{3}$
have a block diagonal form, and while $S=3/2$, $dim[S]=2(\frac{3}{2})+1=4$
, this leads to obtaining the following spin-eigenvalues
\begin{equation}
-3/2\leq m_{S}\leq+3/2,\label{eq:34}
\end{equation}
which means to have

\begin{equation}
m{}_{S}=-\frac{3}{2},\:-\frac{1}{2},\:\frac{1}{2},\:\frac{3}{2},\label{eq:35-1}
\end{equation}
with $S_{\pm}=S_{1}\pm iS_{2}$, $[S_{1},iS_{2}]=2S_{3}$ and $\left\{ S^{2},S_{3},S_{\pm}\right\} $
is a complete set of commuting observables ($CSCO$), and according
to our representation Eq.(\ref{eq:1}), we restrict our self only
to the following eigenvectors 

\begin{equation}
\left|0,\frac{3}{2}\right\rangle ,\left|\frac{3}{2},0\right\rangle ,\left|\frac{1}{2},1\right\rangle ,\left|1,\frac{1}{2}\right\rangle ,\left|0,\frac{1}{2}\right\rangle ,\left|0,\frac{1}{2}\right\rangle .\label{eq:35-2}
\end{equation}

\subsection{{\normalsize{}THE SPIN CURRENT DENSITY}}

We are interested in the spin-eigenvalue $m{}_{S}=\frac{3}{2}$, and
from Eq.(\ref{eq:12}) our wave function with spin is
\begin{equation}
\begin{array}{cc}
\Psi^{\mu\nu}(x,t)=\left(\begin{array}{c}
\psi(x,+\frac{3}{2},t)\\
\chi(x,-\frac{3}{2},t)
\end{array}\right)=\left(\begin{array}{c}
\psi_{\uparrow}\\
\chi_{\downarrow}
\end{array}\right), & \Psi^{\dagger\mu\nu}=\left(\begin{array}{cc}
\psi^{\ast}(x,+\frac{3}{2},t) & \chi^{\ast}(x,-\frac{3}{2},t)\end{array}\right)=\left(\begin{array}{cc}
(\psi_{\uparrow}^{\ast} & \chi_{\downarrow}^{\ast}\end{array}\right)\end{array},\label{eq:35}
\end{equation}
this suggests that the probability density function used to find the
spin-3/2 particle (at a point $x$ and a moment $t$) and also to
determine the current density, in which it is composed of the parts
of the two different spin directions.

The orbital motion of the spin-3/2 particle causes the current density,
where the spin causes a magnetic moment, in which it can be expressed
by its corresponding current. Besides, we note this part of the current
density by $\mathcal{J}_{S}$ (spin current density), and this cannot
occur in the continuity equation because of the indefiniteness of
the total charge.

We determine the spin current density using Maxwell's equation, where
\begin{equation}
rot\overrightarrow{B}=4\pi\left(\mathcal{\overrightarrow{J}}+rot\times\left\langle \mathcal{\overrightarrow{M}}\right\rangle \right),\label{eq:36}
\end{equation}
we have to replace the magnetization $\left\langle \mathcal{\overrightarrow{M}}\right\rangle $
by the average density of the magnetic moment, the up-direction part
of the magnetic dipole is given by
\begin{equation}
\left\langle \mathcal{\overrightarrow{M}}_{\uparrow}\right\rangle =-\mu\psi_{\uparrow}^{\ast}\overrightarrow{S}\psi_{\uparrow},\label{eq:37}
\end{equation}
with
\begin{equation}
\overrightarrow{S}=\frac{1}{2}\sum_{\alpha,\beta=1,2,3}c_{\alpha}^{\dagger}\overrightarrow{\sigma}_{\alpha\beta}c_{\beta},\label{eq:37-3}
\end{equation}
$\overrightarrow{\sigma}_{1},\overrightarrow{\sigma}_{2},\overrightarrow{\sigma}_{3}$
are the usual Pauli matrices, $c_{i}^{\dagger},\:c_{i}$ are creation
and annihilation operators, $\widehat{n}_{i}=c_{i}^{\dagger}c_{i}$
is the number operator of spin-3/2 particles at the site $i$. And
the intrinsic magnetic moment $\overrightarrow{\mu}=\frac{q}{3m}\overrightarrow{S}$
\cite{key-34}, which relates the magnetic moment to the spin angular
momentum, that corresponds to the gyromagnetic ratio $\frac{q}{3m}$
; the highest eigenvalue of the operator $\mu$ is given by
\begin{equation}
\mu_{max}=\frac{\left|q\right|}{2m},\label{eq:37-1-1}
\end{equation}
by the above properties, Eq.(\ref{eq:36}) becomes
\begin{equation}
rot\overrightarrow{B}_{\uparrow}=4\pi\left(\mathcal{\overrightarrow{J}}_{\uparrow}-\mu rot\times\left(\psi_{\uparrow}^{\ast}\overrightarrow{S}\psi_{\uparrow}\right)\right),\label{eq:38}
\end{equation}
with
\begin{equation}
\mathcal{\overrightarrow{J}_{\uparrow}}_{S}=-\mu rot\times\left(\psi_{\uparrow}^{\ast}\overrightarrow{S}\psi_{\uparrow}\right),\label{eq:39}
\end{equation}
knowing that the spin current density is composed of the parts of
the two different spin directions. Noting that for the other different
direction, the spin current density is

\begin{equation}
\mathcal{\overrightarrow{J}_{\downarrow}}_{S}=-\mu rot\times\left(\chi_{\downarrow}^{\ast}\overrightarrow{S}\chi_{\downarrow}\right),\label{eq:43}
\end{equation}
we note that for the up-direction wave function components, we have
only $\{\Psi^{01},\Psi^{02},\Psi^{03}\}$ but for the down-direction
wave function components, we have$\{\Psi^{23},\Psi^{31},\Psi^{12}\}$.

Knowing that the obtained spin current density cannot be affected
by the noncommutativity in space-time, unlike the considered Maxwell's
equation, it would be influenced by the noncommutativity in space-time
as

\begin{equation}
rot\overrightarrow{B}_{\uparrow\downarrow}=4\pi\left(\mathcal{\overrightarrow{J}}_{\uparrow\downarrow}+\mathcal{J}_{nc\uparrow\downarrow}^{2}+rot\times\left\langle \mathcal{\overrightarrow{M}}_{\uparrow\downarrow}\right\rangle \right).\label{eq:41-2}
\end{equation}

\subsection{{\normalsize{}THE HOLSTEIN\textendash PRIMAKOFF TRANSFORMATION}}

As mentioned in the introduction, the graviton with spin-2 in $SUSY$
would have a superpartner called the gravitino with spin-3/2. We want
to show how to connect the spin operators from boson with fermions
spin operators in Fock space using the Holstein\textendash Primakoff
($\mathcal{HP}$) transformation \cite{key-35}, the idea here is
to connect a system of $2N+1$ fermions onto a system of $N$ bosons.
This calculation may be noted in books, and it has been considered
originally by C. E. Alonso and his colleagues in 1992 \cite{key-36}.
The spin and angular momentum eigenfunctions are labeled with the
quantum numbers $n,m,l,j...$

The fermion space is defined in terms of fermion creation (annihilation)
operators $c_{ij}^{+}$$(c_{ij})$. The boson-fermion space is defined
in terms of boson creation (annihilation) operator $B_{ij}^{+}(B_{ij})$,
with new fermion creation (annihilation) operators $a_{ij}^{+}(a_{ij})$,
knowing that the boson and the fermion operators commutate with each
other. With $[B_{ij},B_{\alpha\beta}^{+}]=\delta_{i\alpha}\delta_{j\beta}-\delta_{i\beta}\delta_{j\alpha}$,
$[B_{ij},B_{\alpha\beta}]=0$, the Holstein-Primakoff image of the
single-fermion creation operator $c_{s_{1}m_{1}}^{+}$ is given by
\cite{key-37} 
\begin{equation}
(c_{s_{1}m_{1}}^{+})_{\mathcal{HP}}=\sum_{s_{2}m_{2}}\left\{ \mathcal{T}_{s_{1}m_{1}s_{2}m_{2}}a_{s_{2}m_{2}}^{+}+B_{s_{1}m_{1}s_{2}m_{2}}^{+}a_{s_{2}m_{2}}\right\} ,\label{eq:36-1}
\end{equation}
with the operator $\mathcal{T}=\sqrt{I-B^{+}B}$ that characterizes
the $\mathcal{HP}$ boson expansion, and is to be interpreted by its
Taylor-series expansion, $(c_{ij}^{+})_{\mathcal{HP}}$ denotes the
Holstein-Primakoff image of $c_{ij}^{+}$.

Knowing that the first-order of the expansion of $\mathcal{T}$ (square-root
function) is obtained 

\begin{equation}
\mathcal{T}_{s_{1}m_{1}s_{2}m_{2}}=\delta_{s_{1}s_{2}}\delta_{m_{1}m_{2}}-\frac{1}{2}(B^{+}B)_{s_{1}m_{1}s_{2}m_{2}},\label{eq:44}
\end{equation}
in fact the expansion of $\mathcal{T}$ (with $B^{+}B\ll1$) gives
us $\sum_{i=0}^{\infty}a_{i}(B^{+}B)^{i}$. The Taylor series for
the operator $\mathcal{T}$ is expected to converge very slowly, so
that its development is likely to require many terms, since high order
terms in the series translate into many boson operators. Physically
the utility of such approach is suspicious, so that J. Dukelsky and
S. Pittel \cite{key-38} proposed to accelerate the convergence of
the series by choosing a better initial value for $(B^{+}B){}_{0}$.
This lead to the Eq.(\ref{eq:44}), where 
\begin{equation}
(B^{+}B)_{s_{1}m_{1}s_{2}m_{2}}=\sum_{s_{3}l_{1}l_{2}\alpha\beta}l_{1}l_{2}\alpha(-1)^{s_{3}+m_{1}+l_{2}+\alpha+\beta}X_{j_{2}j_{3}}^{L_{1}}X_{j_{1}j_{3}}^{L_{2}}\left(\begin{array}{ccc}
s_{1} & s_{2} & \alpha\\
m_{1} & -m_{2} & \beta
\end{array}\right)\times\left(\begin{array}{ccc}
l_{1} & l_{2} & \alpha\\
s_{1} & s_{2} & \beta
\end{array}\right)\left(\gamma_{L_{1}}^{+}\times\tilde{\gamma}_{L_{2}}\right)_{\beta}^{\alpha},\label{eq:45}
\end{equation}
we can determine the expression of the boson spin components in the
Fock space
\begin{equation}
\begin{array}{c}
S_{3}=S-n_{b}\\
S^{+}=\sqrt{2s-n_{b}}b\\
S^{-}=b^{+}\sqrt{2s-n_{b}}
\end{array},\label{eq:45-3}
\end{equation}
with $n_{b}=b^{+}b=1,2,3,...$ Knowing that $J=\sqrt{2j+1}$, which
describes the interaction between spins, $b$ is a boson operator
says Holstein-Primakoff. The operator $\gamma_{b}^{+}$ creates the
collective boson of multipolarity $b$ , which is given by

\begin{equation}
\gamma_{\alpha\beta}^{+}=\frac{1}{2}\sum_{j_{1}j_{2}}X_{s_{1}s_{2}}^{\alpha}B_{s_{1}s_{2}\alpha\beta},\label{eq:45-1}
\end{equation}
and $X_{s_{1}s_{2}}^{a}$ are the structure coefficients of $\gamma_{a}^{+}$,
$B_{s_{1}s_{2}\alpha\beta}$ are the angular momentum coupled to boson
creation operators.

To link the quadrupole operator using the $\mathcal{HP}$ expansion,
it is usually possible to directly use the image of the particle-hole
fermion operator, which is given by
\begin{equation}
\left(c_{i}^{+}c_{j}\right)_{HP}=\sum_{k}B_{ik}^{+}B_{jk}+a_{i}^{+}a_{j}.\label{eq:37-1}
\end{equation}

Non-physical states appear for $n_{b}\geq2S$, as under these conditions,
the spin projection on direction 3 is greater than $S$, so that the
use of this spin representation strongly has a meaning for the low
temperature and for large spin values ($n_{b}$ is very weak).

\section{Conclusion}

In conclusion, the space-time noncommutativity is introduced in the
Fisk\textendash Tait equation and subsequently, the continuity equation
is obtained in the cases of commutativity and noncommutativity, without
forgetting that we have investigated the Fisk-Tait equation in NCST
in two cases of potential types. In the first case, we considered
a constant magnetic field, but, for the second case, we considered
a non-constant magnetic field. This, in turn, is responsible for generating
a new quantity (for the first case) which is a correction of density
type. We have shown in the present paper, that the noncommutativity
in space-time affected the continuity equation by causing a new noncommutative
quantity of current density type. In which, this effect does not solve
the problem of the indefiniteness of the total charge, as expected.
This leads us again to think about the difficulties of this type of
equations and its problems that are almost endless. Therefore, the
noncommutativity in space-time is useless in this regard, contrary
to its simplicity and its contribution to solving some problems in
the equations that describe the particles of simple spin such as the
Dirac equation.

Knowing that the space-time noncommutativity effect is introduced
through the Moyal\textendash Weyl product and under the condition
$\Theta=0$, both the Fisk\textendash Tait equation and the continuity
equation in NCST return to that of the usual commutative QM. Furthermore,
we have shown in this work the linking between the fermions and the
bosons in Fock space using the Holstein\textendash Primakoff representation,
without forgetting that we found also the spin current density in
the two different spin directions.

The results of the present work can be used to expand the study of
the motion of the spin-3/2 particles in the NCST. Finally, we point
out that we are out looking to investigate the reflexion and the transmission
coefficients depending on these results to investigate the Klein paradox.
\begin{acknowledgments}
The author would like to thank Pr Lyazid Chetouani for his interesting
comments and suggestions, and appreciates the unknown referee\textquoteright s
valuable and profound comments.\end{acknowledgments}

\end{document}